# MUON COOLING

E.G. Bessonov, M.V. Gorbunkov, Lebedev Phys. Inst. RAS, Moscow, Russia,
A.A. Mikhailichenko, Cornell University, Ithaca, NY 14853, U.S.A.

*Abstract*

The possibility of the Enhanced Optical Cooling (EOC) of muon beams in storage rings is investigated.

## INTRODUCTION

Muon in motion has a lifetime $\tau_\mu = \tau_{\mu,0} \gamma$, where $\tau_{\mu,0} \cong 2.2 \mu s$ is the muon lifetime at rest, $\gamma = \varepsilon / m_{\mu,0} c^2$, $\varepsilon$ stands for muon energy. During this time, muons can develop $N_\mu |_{\gamma \gg 1} = \tau_\mu c / C \cong 297 \overline{B}(T)$ revolutions in a storage ring having the orbit circumference $C$ and average magnetic field strength $\overline{B}$. Damping time of the muon beam in a storage ring must be much smaller than the muon decay time: $\tau < \tau_\mu$. This is the most severe limitation on the muon cooling in storage rings. Enhanced (fast) Optical Cooling methods must be used in this case.

Below we investigate the possibility of muon cooling in a storage ring using version 1, section 3 of EOC [1].

## COOLING SCHEME

Stochastic cooling (SC) and optical stochastic cooling (OSC) of particle beams in rings were considered in [2] - [4]. Our scheme is close to the OSC one. It includes a storage ring, pickup and kicker undulators installed in straight sections of the ring, an optical system which includes an optical filter, an optical parametric amplifier (OPA) [1]. It also includes a procedure and adequate hardware for selection of undulator radiation (UR) wavelets (URW) emitted by particles in the pickup undulator. This selection is arranged by means of a screen installed at the image plane of the optical system connecting the pickup and kicker undulators.

Maximal rate of energy loss for a particle in the combined fields of one helical kicker undulator and URW amplified in OPA is

$$P_{loss}^{max} = -eE_w^{cl} L_u \beta_{\perp m} f \, \Phi(N_{ph}^{cl}) \sqrt{\alpha_{ampl}} \text{ or}$$

$$P_{loss}^{max} = \frac{2\sqrt{2} \pi e^2 \gamma f \, K^2 \sqrt{M} \Phi(N_{ph}^{cl}) N_{kick} \sqrt{\alpha_{ampl}}}{(1+K^2)^{3/2} \sigma_w}, \quad (1)$$

where $E_w^{cl} = \sqrt{2} r_\mu \gamma^2 \sqrt{\overline{B^2}} / (1+K^2)^{3/2} \sqrt{M} \sigma_w$ is the electric field strength related to the first harmonic of the undulator radiation emitted by a particle in the pickup undulator within the frequency band $\Delta \omega / \omega = 1/2M$ calculated in the framework of the classical electrodynamics (CED), $K = e \sqrt{\overline{B^2}} \lambda_u / 2\pi m_\mu c^2$, $\sqrt{\overline{B^2}}$ is the undulator r.m.s. magnetic field strength, $\beta_{\perp m} = K / \gamma$, $\Phi(N_{ph}^{cl})|_{N_{ph}^{cl} < 1} = \sqrt{N_{ph}^{cl}}$, $N_{ph}^{cl} = \pi \alpha K^2 / (1+K^2)$ is the number of photons emitted by one particle in the URW, $M$ is the number of the undulator periods, $L_u = M \lambda_u$, $\lambda_u$ is the undulator period, $\alpha_{ampl}$ is the gain in OPA, $f$ is the revolution frequency, $\sigma_w$ is the waist size of the URW. Function $\Phi(N_{ph}^{cl})$ takes into account the quantum nature of the particle emission of the electromagnetic radiation. It radiates on average one photon per $1/N_{ph}^{cl}$ pass throw pickup undulator in the photon energy interval $\Delta(\hbar\omega)/\hbar\omega = 1/2M$ near the maximum photon energy $\hbar\omega_{max}$. Contrary, in this case the electric field strength and the stimulated energy transfer in the kicker undulator are only $1/\sqrt{N_{ph}^{cl}}$ times bigger than the ones calculated in the framework of CED.

We suggested that the density of the energy in the URW emitted by a particle, is approximated by Gaussian distribution with a waist size $\sigma_w > \sigma_{x,z}^\mu$, $\sigma_{w,c}$ within the length $2M\lambda_{1\min}$, where $\sigma_{x,z}^\mu$, are the transverse beam dimensions; a value $\sigma_{w,c} = \sqrt{L_u \lambda_{1\min}/8\pi}$ is the waist longitudinal dimension corresponding to $Z_R = L_u/2$; the Rayleigh length $Z_R = 4\pi\sigma_w^2/\lambda_{1\min}$, $\lambda_{1\min} = \lambda_1|_{\theta=0}$, $\lambda_1 = \lambda_u(1+K^2+\vartheta^2)/2\gamma^2$ is the wavelength of the first harmonic of the UR emitted at the angle $\theta$, counted between the vector of particle average velocity in the undulator and the direction to the observation point, $\vartheta = \gamma\theta$.

The damping times for the particle beam in the longitudinal and transverse planes are [1]

$$\tau_\varepsilon = \frac{6\sigma_{\varepsilon,0}}{P_{loss}^{max} \cdot N_k} = \frac{3\sqrt{2}(1+K^2)^{3/2} T \sigma_w \sigma_{\varepsilon,0}}{2\pi r_e m c^2 \gamma K^2 \sqrt{M} \Phi(N_{ph}^{cl}) N_{kick} \sqrt{\alpha_{ampl}}},$$

$$\tau_x = \tau_\varepsilon \frac{\sigma_{x,0}}{\sigma_{x_\eta,0}}, \quad (2)$$

where $\sigma_{\varepsilon,0}$ is the initial energy spread of the particle beam, $\sigma_{x_\eta,0} = \eta_{x,k} \beta^{-2}(\sigma_{\varepsilon,0}/\varepsilon_s)$, $\sigma_{x,0} = \sqrt{\beta_{x,k} \in_x}$ are the initial radial beam size in kicker undulator determined by the energy spread and the spread of betatron amplitudes, $\eta_{x,k}$, $\beta_{x,k}$ are the ring dispersion and beta functions at the kicker undulator, $\in_{x,0}$ is the initial radial emittance of the beam, $N_k$ is the number of kicker undulators, $T = 1/f$, the product $r_e m_e c^2 = e^2$ of the electron radius and mass are introduced for the convenience. Note, that the damping time for the transverse direction is proportional to $\beta_{x,k}/\eta_{x,k}$. Factor 6 in (2) takes into account a circumstance that the initial energy spread is $2\sigma_{\varepsilon,0}$, the particles meet with their URWs 2 times less often (half of

URWs are absorbed by the screen) and that the jumps of the particle closed orbit in average lead to lesser jumps of the amplitudes of synchrotron and betatron oscillations. The path between pickup and kicker undulators must be isochronous to the fraction of operational wavelength.

When $N_{ph}^{cl} < 1$, $N_{mix} = N_k$, the equilibrium energy spread and the beam dimensions at the location of kicker undulator are

$$\sigma_{\varepsilon,\,eq} = \frac{N_{ph,\Sigma}}{2\sqrt{2}N_{ph}^{cl}} \Delta\varepsilon_{loss}^{max}, \quad \sigma_{x,\,eq} = \sigma_{x_\eta,\,eq} = \frac{\eta_{x,k}}{\beta^2}\frac{\sigma_{\varepsilon,\,eq}}{\varepsilon}. \quad (3)$$

where $N_{ph,\Sigma} \cong N_{ph}^{cl}(N_{p,s}-1) + N_n > 1$ is the number of photons in the sample produced by $N_{p,s}-1$ extraneous particles and $N_n$ noise photons at the amplifier front end, $N_{p,s} = 2M\lambda_{1,min}N_p/\sigma_{s,0}$ is the number of particles in the URW sample, $N_p$ stands for the number of particles in the bunch, $\Delta\varepsilon_{loss}^{max} = P_{loss}^{max}/f N_{ph}^{cl}$ is the particle energy jump in the fields of one kicker undulator and the amplified URW (if one photon of the energy $\hbar\omega_{1,max}$ in the URW), $N_{mix}$ is the number of mixers in the ring, $\sigma_{s,0}$ is the initial length of the particle bunch. The energy jump $\Delta\varepsilon_{loss}^{max}$ corresponds to the energy loss by each particle in the pickup undulator if URW contains a single photon only. Note, that if the number of noise photons in the equation (3) set to $N_n = 0$, then $N_{ph}^{cl}$ in the equation will disappear as well. It means that classical and quantum expressions for equilibrium beam dimensions in this case are identical (differences stay only in the energy jumps $\Delta\varepsilon_{loss}^{max}$ for classical and quantum cases). For one particle the equilibrium beam dimensions would be limited only by one energy jump.

The equations (3) determine the particle energy jump through the equilibrium energy spread and dimensions of the beam. Below we introduce the dimensionless coefficients $k_\varepsilon = \sigma_{\varepsilon,0}/\sigma_{\varepsilon,eq}^{EOC} > 1$, $k_x = \sigma_{x,0}/\sigma_{x,eq}^{EOC} > 1$. The values $k_{\varepsilon,x} - 1$ present the degree of cooling for the longitudinal and transverse directions respectively. In this case the energy jump (3) corresponding to the equilibrium beam parameters can be represented in the form $\Delta\varepsilon_{loss,\varepsilon,x}^{max} = 2\sqrt{2}N_{ph}^{cl}\sigma_{\varepsilon,0}/k_{\varepsilon,x}N_{ph,\Sigma}$. The corresponding power loss is

$$P_{loss,\varepsilon,x}^{max} = f N_{ph}^{cl}\Delta\varepsilon_{loss,\varepsilon,x}^{max} = \frac{2\sqrt{2}f(N_{ph}^{cl})^2\sigma_{\varepsilon,0}}{k_{\varepsilon,x}N_{ph,\Sigma}}. \quad (4)$$

We took into account that from the equation $\sigma_{x,\,eq} = \sigma_{x_\eta,\,eq}$ follows the equation $P_{loss,x}^{max} = P_{loss,\varepsilon}^{max}$.

In this case, according to (2), the damping time (the time necessary to decrease the initial beam dimensions to the equilibrium ones) is

$$\tau_{\varepsilon,x} = \frac{6\sigma_{\varepsilon,0}}{P_{loss,1}^{max}\cdot N_k} = \frac{3\sqrt{2}N_{ph,\Sigma}k_{\varepsilon,x}}{2(N_{ph}^{cl})^2 N_k}T. \quad (5)$$

According to the expression (5), the damping time is proportional to the degree of cooling, to the revolution period over the number of kicker undulators and to the number of noise photons at the amplifier front end plus the number of muons in the URW sample.

If the degree of cooling and the equilibrium beam parameters are chosen, then according to (1) and the condition $P_{loss}^{max} = P_{loss,\varepsilon,x}^{max}$, the gain corresponding to the equilibrium beam parameters is

$$\sqrt{\alpha_{ampl}} = \frac{(1+K^2)^{3/2}(N_{ph}^{cl})^2\sigma_w}{\pi r_e m_e c^2\gamma K^2\sqrt{M}\Phi(N_{ph}^{cl})k_{\varepsilon,x}N_{ph,\Sigma}}\sigma_{\varepsilon,eq}. \quad (6)$$

It follows from (5) that one can set an arbitrary degree of cooling if the gain of the OPA and equilibrium dimensions of the beam are chosen according to (6). Higher initial beam dimensions require higher gain in OPA, the rate of cooling ($\Delta\varepsilon_{loss}^{max}$) and the equilibrium beam dimensions. In the case of fast muon cooling the conditions could be very hard to satisfy in practice.

If we set the damping time equal to the muon decay time ($\tau_\mu = \tau_{\varepsilon,x}$) then the equation (5) determines the relativistic factor of the being cooled muons for this case:

$$\gamma_\mu = \frac{\tau_\mu}{\tau_{\mu,0}} = \frac{\tau_{\varepsilon,x}}{\tau_{\mu,0}} = 4.5\cdot 10^5 \tau_{\varepsilon,x}. \quad (7)$$

The deflection parameter of the undulator is determined by the given values $\gamma_\mu$, $\lambda_{1,min}$ and $\lambda_u$:

$$K_1^2 = 1 + \frac{2\gamma_\mu^2\lambda_{1,min}}{\lambda_u}. \quad (8)$$

The average power of an optical amplifier is determined by the power of its URW beam and by its noise power:

$$P_{ampl} = P_{ampl}^{URW} + P_{ampl}^n, \quad (9)$$

where $P_{ampl}^{URW} = \hbar\omega_{1,max}N_{ph}^{cl}\alpha_{ampl}fN_\mu N_b/2$, $P_{ampl}^n = l_b\alpha_{ampl}\hbar\omega_{1,max}fN_nN_b/2M\lambda_{1min}$, $\omega_{1,max} = 2\pi c/\lambda_{1,max}$, $l_b$ is the length of the particle bunch and $N_b$ is the number of bunches. The powers correspond to the case in which a half of total number of particles are involved in the cooling process (screening is introduced) and the amplification time interval of the amplifier is equal to the duration time of the particle bunch [1].

The transverse resolution of the particle within the bunch is

$$\delta x_{res} \cong 0.86\sqrt{\lambda_{1min}L_u}. \quad (10)$$

To have the smallest possible damping time, we must choose small beta-function value in vicinity of kicker undulators (limited by condition $\beta_{x,k} > L_u/2$) to have smallest transverse beam dimension in a damping ring.

The number of revolutions performed by muon in a ring during one period of phase oscillations comes to

$$N_{rev}^{ph} = \sqrt{2\pi\varepsilon_s/h\alpha_c eV_m}, \quad (11)$$

where $h$ is the harmonic number, $\alpha_c$ is the momentum compaction factor, $V_m$ is amplitude of the RF accelerating voltage.

## EXAMPLE

To appreciate the efficiency of the EOC, we will consider below an example of the EOC of muon beam in a storage ring with the equal number of pickup and kicker undulators, mixers, OPAs, isochronous bends between pickup and kicker undulators and non-isochronous residual part of the orbit. Let the revolution period be $T = 15 \mu s$, the undulator period $\lambda_u = 1$ m, the number of the undulator periods M=10, the number of undulators $N_p = N_k = 10$, limiting degree of cooling $k_{\varepsilon,x} - 1 = 10$, number of muons per a bunch $N_\mu = 10^7$, the number of bunches $N_b = 10^2$, the length of the muon bunch $l_b = 2$ m, the wavelength of the OPA $\lambda_{1\min} = 4 \cdot 10^{-5}$ cm, the waist size of the URW at kicker undulator $\sigma_w = 1$ mm, the number of noise photons at the amplifier front end $N_n = 1$, the equilibrium beam energy spread $\sigma_{\varepsilon,eq} = 1.84 \cdot 10^6$ eV ($\sigma_{\varepsilon,eq}/\varepsilon = 10^{-6}$), the equilibrium transverse dimensions of the beam at kicker undulator $\sigma_{x_\eta,k\,eq} = \sigma_{x,k,eq} = 0.3$ mm.

In this case, according to (5)-(8), the muon decay time $\tau_\mu = 38.1$ ms, the muon energy $\varepsilon_\mu \cong 1.836$ TeV ($\gamma_\mu = 17.4 \cdot 10^3$), the deflection parameter of the undulator $K = 22$, the gain of OPA $\alpha_{ampl} = 5.13 \cdot 10^{10}$, the undulator magnetic field strength $\sqrt{B^2} = 48.7$ T, the initial energy spread $\sigma_{\varepsilon,0}/\varepsilon = 10^{-5}$, the photon energy $\hbar\omega_{1,\max} = 3.1$ eV, $\sigma_{w,c} = 0.378$ mm, $\delta x_{res} \cong 2$ mm, $P^{URW}_{ampl} = 1.76$ kW, $P^n_{ampl} = 3.82$ W, $P_{ampl} = 5.58$ kW, $T_\varphi = 21.8$ ms.

If we choose the dispersion functions of the ring at pickup and kicker undulators equal to $\eta_{x,p} = 10^3$ m, $\eta_{x,k} = 30$ m, the beta functions $\beta_{x,p} = 50$ m, $\beta_{x,k} = 5$ m, then the transverse initial dimensions of the beam at the pickup undulator $\sigma_{x_\eta,p,0} = 10$ mm, $\sigma_{x,p,0} = 10$ mm, the total transverse initial dimensions of the beam at the pickup undulator $\sigma_{\Sigma,p,0} = \sqrt{\sigma^2_{x_\eta,p,0} + \sigma^2_{x,p,0}} = 14$ mm (amplitudes will be $\sqrt{2}$ times larger).

We took into account that $N^{cl}_{ph}|_{K \gg 1} \simeq \pi\alpha = 2.3 \cdot 10^{-2}$, $N_{\mu,s} = 40$, $N_{ph,\Sigma} = 1.9$, $1/N^{cl}_{ph} = 43.6$, $\Phi(N^{cl}_{ph}) = 0.151$.

## DISCUSSION

Short muon lifetime together with presence of noise at the amplifier front end, quantum nature of the emission of URWs (small probability of photon emission by muon in pickup undulators) in all schemes of optical cooling impose severe limitations on the minimal rate of cooling, OPA gain, its average power, on the number of muons being cooled. High rate of cooling at moderate initial beam energy spread and transverse beam dimensions leads to high muon energy in a damping ring, big jumps of muon energy, position of closed orbit and amplitude of betatron oscillation. It leads also to big values for equilibrium beam dimensions at modest degree of cooling $k_{\varepsilon,x} \sim (10)$, high power of OPAs at small number of particles in the sample and hence in the bunch, high magnetic field strength (high temperature SC magnets with~50T field for undulators to shorten their length could be used here). High length of pickup undulator leads to low resolution of particles in the beam necessary for the EOC scheme and to high dispersion and beta functions in the location of the pickup undulator.

Note that quantum nature of URW emission at $N^{cl}_{ph} \ll 1$ leads to small number $N^{cl}_{ph} \ll 1$ of emitted URWs per a turn and to fewer ($\sim \sqrt{1/N^{cl}_{ph}}$) degree times increase of the energy transfer in the process of stimulated interaction of the URW with the muon at kicker undulator (reduction factor $\Phi(N^{cl}_{ph}) \sim 0.1$) [1].

In the EOC scheme a non-exponential cooling regime is realized when the beam dimensions are decreased much more than $e=2.7$ times for one damping time.

According to (5), (7), the damping time and the energy of the beam necessary for cooling at given revolution period are proportional to the number of the being cooled particles (number of particles in the sample) and the degree of cooling. In our example the degree of cooling $k_{\varepsilon,x} - 1 = 10$ and at the number of particles in the sample $N_{\mu,s} = 40$ lead to the muon beam energy $\varepsilon_\mu \cong 1.836$ TeV and the muon beam damping time is $\tau_\mu = 38.1$ ms, which is equal to the muon lifetime at this energy. If the phase advance between pickup and kicker undulators is equal to $\pi$ then the beam is cooled both in longitudinal and transverse dimensions simultaneously. The degree of cooling of the 4-dimensional phase space is $10^4$. If the screen in the optical system is transparent, produce phase advance $\pi$ for the URW and overlaps half of the beam in the image plane along the position of the synchronous orbit then all particles both with positive and negative deviations of their energy from synchronous one will take part in the cooling process. In this case the damping time will be 2 times smaller.

## SUMMARY

We have demonstrated the possibility of EOC of muons in storage rings. In contrast to the publication [4] we took into account the quantum nature of the photon emission in the pickup undulator. According to (1) and (5), (7) it leads to the decrease of the rate of cooling by the coefficient $\Phi(N^{cl}_{ph}) = 0.15$ for each degree of freedom (longitudinal and horizontal transverse) and to the increase of the energy necessary for cooling.

The rate of the energy loss by muons for the EOC scheme stays constant. It does not depend on the deviation of their energy from the energy of given reference muon. In this case the smaller initial maximal transverse and longitudinal amplitudes of oscillations of particles in the beam the smaller the equilibrium beam dimensions at the same damping time could be reached.

Despite what we underlined in our previous publication [1] the importance of quantum effects in case when the number of radiated photons is less than one, (or in other words–the energy radiated, when calculated by classical formula, is less than the energy of quanta) this fact has not attracted adequate attention. All publications of other authors on optical cooling (including recent one [5]) need to be corrected for the rate of cooling which actually is ~15% only of what they are claiming for each degree of freedom. The equilibrium beam dimensions must be corrected as well. We would like to attract attention once again that these limitations manifest themselves at the *kicker* undulator, as the stimulated processes here are proportional to the *electric field* value in a wavelet (not to its energy). Additional remarks one can find in the Appendix.

EOC and other schemes of OSC can be used effectively for cooling of stable particles (protons, ions). EOC scheme has faster non-exponential rate of cooling but it requires optics with high resolution and OPAs with shorter wavelengths.

The possibility of EOC and other schemes of OSC for muons in storage rings requires additional investigations.

This work was supported by RFBR under Grant No 09-02-00638-a.

## APPENDIX

In external fields of bending magnets and undulators installed in the storage rings, relativistic particles emit electromagnetic radiation in the form of synchrotron (SR) and UR wavelets directed in the narrow range of angles ($\sim 1/\gamma$) along their velocities. The spectral-angular distribution of the emitted energy and polarization characteristics of the URWs can be calculated in the framework of classical electrodynamics (CED). More correct quantum approach leads to the same characteristics of the emitted radiation if we accept that the energy is emitted in the form of photons with the energy $\varepsilon_{ph} = \hbar\omega$ and if the energy of these photons is much less than the energy of the particle. Distribution of the number of photons emitted by the particle in the bending magnet or in the undulator is determined by the distribution of the energy of the emitted radiation

$$\frac{\partial N_{ph}}{\partial \omega \partial o} = \frac{1}{\hbar\omega} \frac{\partial^2 \varepsilon}{\partial \omega \partial o} .$$

The forms of the SR and UR wavelets and their properties are different. The loss of the energy (friction) leads to damping of betatron and phase oscillations of particles in the rings. In CED the energy is emitted continuously. In the quantum electrodynamics (QED) particles propagate along trajectory some distance without the energy loss and then emit instantaneously the URW with the energy $\hbar\omega$ and at the same time they loose the energy $\Delta\varepsilon = \hbar\omega$. The radiation in the QED has random nature. In average the energy emitted in the framework of CED and QED are equal. The emission of the energy in the form of finite quanta leads to jumps of the closed orbit of the particles. Joint action of the quanta excitation and damping leads to rms equilibrium transverse and longitudinal dimensions of particle beams in the rings. The value of these dimensions is proportional to the value of the jumps (the energy of URWs) and in strict accordance with the experiment. The URW length, special and time dependence, polarization and other characteristics of radiation are determined by the type of undulators and in agreement with the experiment.

Note that stable particles exist in the ring for many hours. It is possible to inject in the ring one particle and produce experiments [6]. Experimentally we can detect the particle (reduction of the particle state), observe its Brownian motion in the equilibrium volume. If we use laser undulator then we can have very hard radiation and lose the particle if the jump of the particle orbit will be larger than the dynamic aperture of the ring [7]. URW consists of a whole number of photons (never parts of photons).